\documentclass[aps,prl,twocolumn,superscriptaddress]{revtex4}
\usepackage{graphicx}
\usepackage{mathrsfs}

\begin{document}



\title{Fabrication of large addition energy quantum dots in graphene}

\author{J. Moser}
\email{moser.joel@gmail.com} \affiliation{CIN2 (CSIC-ICN) Barcelona,
Campus UAB, E-08193 Bellaterra, Spain\\}

\author{A. Bachtold}
\email{adrian.bachtold@cin2.es} \affiliation{CIN2 (CSIC-ICN)
Barcelona, Campus UAB, E-08193 Bellaterra, Spain\\}



\begin{abstract}
We present a simple technique to fabricate graphene quantum dots in
a cryostat. It relies upon the controlled rupture of a suspended
graphene sheet subjected to the application of a large electron
current. This results in the {\it in-situ} formation of a clean and
ultra-narrow constriction, which hosts one quantum dot, and
occasionally a few quantum dots in series. Conductance spectroscopy
indicates that individual quantum dots can possess an addition
energy as large as $180\;$meV. Our technique has several assets: (i)
the dot is suspended, thus the electrostatic influence of the
substrate is reduced, and (ii) contamination is minimized, since the
edges of the dot have only been exposed to the vacuum in the
cryostat.
\end{abstract}


\maketitle

Graphene can be seen as a giant, flat molecule whose shape can be
tailored by means of standard fabrication techniques. One long term
goal is to structure graphene down to a small molecule, such as a
benzene ring, connected to two graphene electrodes. A significant
step in this direction has been achieved with the fabrication of
chains of carbon atoms using the electron beam of an electron
microscope to knock off the atoms of a graphene sheet \cite{jin}.
Top-down fabrication strategies to structure graphene represent an
original approach to realize devices in molecular electronics.
Indeed, studying charge transport across molecules has so far been
relying on a bottom-up approach, whereby the molecule under study is
typically synthesized via chemical means and later on placed between
two large metal electrodes. In this configuration, the contacted
molecule often acts as a quantum dot, and the addition energy
$E_{\rm add}$ needed to add one electron has been measured to be
100-400~meV \cite{kubatkin,osorio,natelson,wernsdorfer}.

\begin{figure}[t]
\includegraphics[width=5cm]{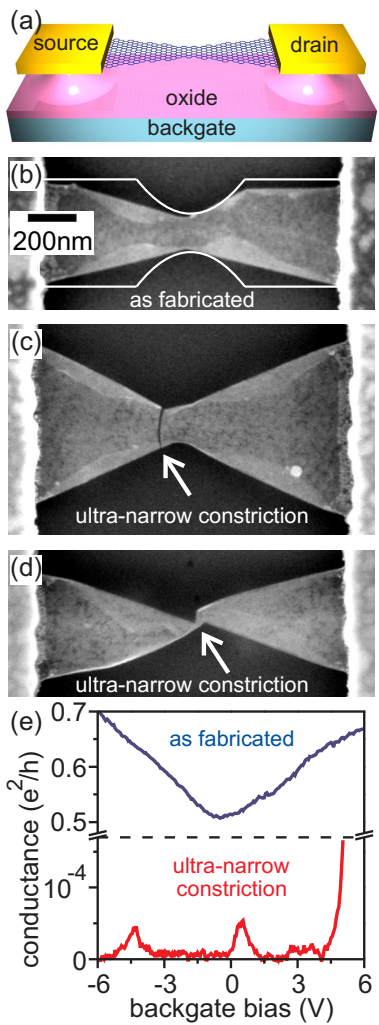}
\caption{(color) (a) Schematic of our device, showing a suspended
graphene sheet contacted by source and drain electrodes. (b)
Scanning electron microscope image of the device (top view) before
rupture of the graphene sheet. The white contour indicates the shape
of the sheet measured by atomic force microscopy prior to oxide
removal. The edges of the sheet fold during fabrication of the
suspended device. (c), (d) A large current induces a rupture of the
sheet (arrow), shown here for two samples. In (d), the sheet is
shown to break and then fold. (e) Two-point conductance vs. backgate
bias at $T=10$~K, before (blue trace) and after (red trace) breaking
of the sheet. Note the two conductance scales. For this device,
conductance oscillations are not periodic, and the stability diagram
(not shown) suggests the presence of SET's in series.}
\end{figure}

In this Letter, we present a technique to fabricate quantum dots out
of a graphene sheet, under high vacuum and at low temperature in a
cryostat. It is based on the controlled rupture of a suspended
graphene sheet subjected to a large electron current, resulting in
the formation of a quantum dot with an addition energy as large as
180~meV. Even though this energy is large, simple estimates show
that the size of the quantum dot is of the order of 10~nm, that is
the size of a large molecule.

\begin{figure}[t]
\includegraphics[width=5cm]{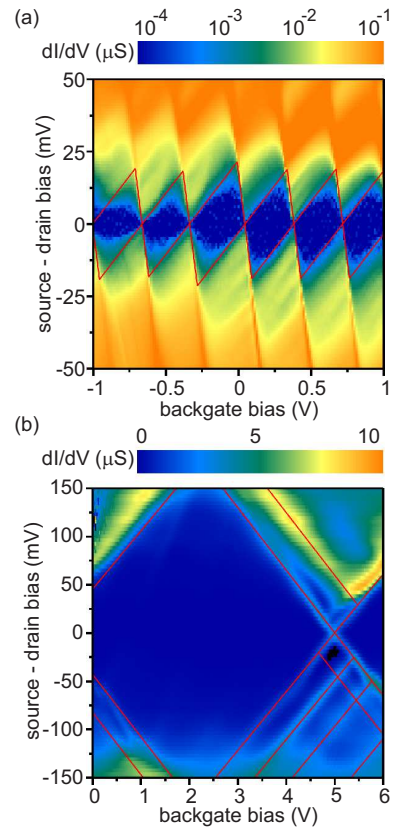}
\caption{(color) Differential conductance $dI/dV$ as a function of
source-drain bias $V$ and backgate bias for two samples (a), (b) at
$T=10$~K. Red lines are guides to the eye.}
\end{figure}

We start with a description of our fabrication process. Our
substrates are highly doped silicon wafers coated with $440\;$nm of
thermal silicon oxide. We thoroughly clean the oxide surface in
ozone in order to remove adsorbed hydrocarbon chains. Shortly
afterwards, we deposit graphene sheets on the oxide surface using
the scotch tape technique \cite{novoselov}. We identify single layer
graphene sheets among thicker flakes by measuring the reflected
light intensity using the blue channel of a CCD camera mounted on a
Nikon optical microscope. We reduce the width of the middle region
of the graphene sheet down to $\sim 200\;$nm using electron beam
lithography (EBL) followed by reactive ion etching in oxygen (see
white contour in Fig.~1(b)). We pattern source and drain electrodes
by EBL and Cr/Au thermal deposition. At this point, we remove
contamination by annealing the device at $300^\circ$C in flowing
Ar/H$_2$ for a few hours. We then suspend the graphene sheet by
removing the oxide in an aqueous solution of hydrofluoric acid
(Fig.~1(a)). We expect that, by this stage of the process, only
those oxide regions that are masked by graphene sheets have remained
hydrocarbon-free (owing to the ozone treatment), and thus
hydrophilic. As a result, the etching solution can rapidly diffuse
between graphene and the oxide surface. This guarantees that even
large graphene sheets are freely suspended. The oxide etched away,
we critical point dry the device. During this fabrication step, the
sheet is always observed to fold on each side of the constriction
(compare white contour before wet etching with the actual shape of
the sheet in Fig.~1(b)).

At $T=15\;$K and under high vacuum ($<10^{-6}$~mbar), we employ the
current-induced cleaning technique \cite{cleaning} to remove
adsorbates off the surface of the graphene sheet, which reduces
extrinsic doping. A two-point measurement of the conductance as a
function of the bias applied to the silicon substrate (backgate bias
$V_{g}$) shows a conductance minimum close to zero $V_{g}$ for some
devices (top curve in Fig.~1(e)).

Passing a larger current between the source and drain electrodes
drives the device to near mechanical rupture and forms an
ultra-narrow constriction (see arrow in Figs.~1(c), 1(d)). Depending
on the width of the patterned middle region, rupture is observed for
currents ranging from 100 to $200\;\mu$A, consistent with the
typical breakthrough current of graphene of about
$1\;\mu\rm{A}/\rm{nm}$ \cite{amelia}. The rupture occurs in the
vicinity of the patterned middle region: the reduced cross-section
ensures that the current density is largest, which favors the
rupture. Upon reaching maximum current, the conductance exhibits
large jumps \cite{bockrath,tour}. As soon as the conductance falls
to a fraction of its low current value, a computer controlled
feedback loop takes the current back to zero within $\sim10$~msec.
The lower conductance is consistent with the formation of a narrower
constriction.

Following this large current treatment, we find that the low
source-drain bias conductance $G$ of the graphene sheet oscillates
as a function of backgate bias $V_{g}$ (Fig.~1(e)). Measuring the
differential conductance $dI/dV$ as a function of source-drain bias
$V$ and $V_{g}$ can yield well defined Coulomb diamonds for some
devices, which indicate the presence of electron charging islands
\cite{weis1} in the ultra-narrow constriction. Figs.~2(a) and 2(b)
display typical examples of such Coulomb diamonds, measured in two
different samples at $T=10\;$K. In Fig.~2(a), the almost constant
width of the diamonds (along the backgate bias axis), their constant
slopes (highlighted by red lines), and the fact that diamonds are
closed, all indicate the presence of one single electron island in
the ultra-narrow constriction. In the Coulomb blockade regime, the
diamonds height along the $V$-axis is a measure of the energy
$E_{\rm add}$ to add one charge carrier to the island. In Fig.~2(a),
$E_{\rm add}\simeq20\;$meV. In Fig.~2(b), we measure a strikingly
large $E_{\rm add}$ of $\sim180$~meV. This value is nearly one order
of magnitude larger than $E_{\rm add}$ in graphene single electron
transistors nanofabricated so far
\cite{ponomarenko,schnez,johannes,todd,stampfer}. Our fabrication
technique has a reasonable yield: out of 7 devices studied, 3
exhibited single-electron transistor (SET) behavior over the range
of backgate biases explored, 2 showed signatures of a series of
SET's, 1 exhibited a particularly large $I(V)$ gap of $\sim700$~mV
at $T=10\;$K (which at present is not understood), and 1 failed
during the high current treatment \cite{note}.

Fig. 2(b) shows additional conduction channels that appear at larger
$V$ as $dI/dV$ resonances running parallel to the diamond edges.
Fig.~3 shows $dI/dV$ curves as a function of $V_{g}$ at various
values of $V$ and at a bath temperature of 50~mK for the same
sample. The energy spacing between consecutive conduction channels,
given by $V$ at the onset of a $dI/dV$ resonance, is $\delta E\simeq
25$~meV \cite{thermal}. These resonances are suggestive of transport
mediated by excited states in the dot. Indeed, strong spatial
confinement gives rise to a spectrum of zero-dimensional (0-D)
levels that host excited states above the lowest available quantum
level (ground state) \cite{johnson,weis2}. These 0-D excited states
open up additional transport channels in the non-linear regime
(larger $V$), and have been clearly observed in graphene quantum
dots \cite{schnez,johannes}.

\begin{figure}[t]
\includegraphics[width=5cm]{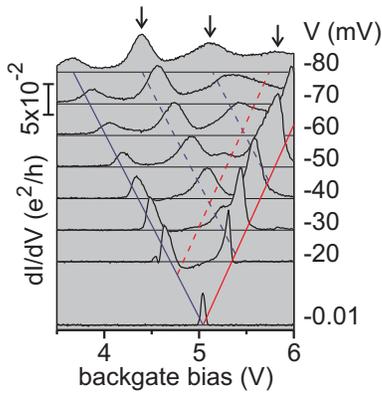}
\caption{(color) Differential conductance $dI/dV$ traces, offset for
clarity, as a function of backgate bias and at various source-drain
biases $V$ at $T=50$~mK for the sample of Fig.~2(b). $dI/dV$
resonances indicated by arrows are presumably related to excited
states.}
\end{figure}

We can obtain a rough estimate for the size $d$ of our quantum dot
in Fig.~2(b) using the measured addition energy $E_{\rm add}$. Using
the disk model, $d={\rm{e}}^2/(4\epsilon_{0}E_{\rm{C}})\simeq
24$~nm, where $\epsilon_{0}$ is the vacuum permittivity, and
assuming that the charging energy $E_{\rm C}\simeq E_{\rm add}$.
Another estimate can be obtained by comparing $E_{\rm add}$ to
values measured by others for graphene dots of various sizes,
assuming similar capacitive couplings. Ponomarenko and coworkers
\cite{ponomarenko} studied several graphene quantum dots whose
diameters range from 40~nm to 250~nm. When we plot their measured
$E_{\rm add}$ as a function of dot diameter $d$, we find that
$E_{\rm add}$ scales roughly as $E_{\rm add}\simeq 500
\rm{meV}\cdot\rm{nm}/d$. Assuming that $E_{\rm add}$ is
predominantly given by the charging energy $E_{\rm{C}}$ and taking
into account that the average dielectric constant felt by the
suspended graphene sheet in our configuration is about twice
smaller, we estimate that the diameter of our quantum dot is $\sim
6$~nm. Overall, our dot size is on the $\sim 10$~nm scale. Finally,
we estimate the number $N$ of charge carriers in the graphene dot
assuming a disk of diameter $d=10\;$nm and provided that $\delta
E=25\;$meV, which reads $N=(\hbar\rm{v_{F}}/d\delta{E})^2\simeq 10$
\cite{schnez}. Surprisingly, this indicates a very large charge
density of $\sim 10^{13}/\rm{cm}^2$. Such a large density may
originate from chemical doping by molecules \cite{water} (still
present under high vacuum) that have interacted with the dangling
bonds at the edges of the dot.

We now comment on a possible mechanism driving the formation of our
quantum dots. Due to its finite width, the ultra-narrow constriction
created by the current-induced rupture of the graphene sheet may act
as a hard wall confinement potential for charge carriers that opens
an energy gap at the Dirac point, as observed in graphene
nano-ribbons \cite{kim,stampfer}. Following Todd, \textit{et al.}
\cite{todd}, and Stampfer, \textit{et al.} \cite{stampfer}, we
propose that charge carriers become localized by potential
fluctuations along the ultra-narrow constriction. These fluctuations
may originate from the molecules \cite{water} having reacted with
the dangling bonds at the edges of the dot. Fig.~4 illustrates this
scenario, where a fluctuating potential defines 0-D states above the
energy gap created by the ultra-narrow constriction.

\begin{figure}[t]
\includegraphics[width=5cm]{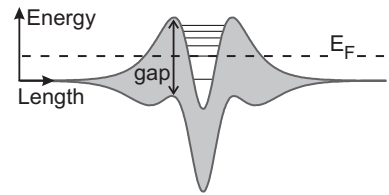}
\caption{Proposed potential landscape giving rise to 0-D
confinement. The ultra-narrow constriction opens a gap at the Dirac
point. On top of this gap, a fluctuating potential defines the
quantum dot.}
\end{figure}

The mechanical robustness of our suspended devices is also
noteworthy. Because graphene sheets, in every likelihood, contain
finite build-in strain, it is remarkable that our tiny dots, which
are only weakly connected to two sections of free standing graphene
sheets, can exist at all. Presumably, the folding of the sheet along
the edges make the graphene sheet stiffer (Fig.~1(b)), and the
overall device mechanically more stable.

In conclusion, we have shown that the rupture of a graphene sheet
subjected to a large current can be harnessed to fabricate graphene
quantum dots endowed with a large addition energy. The fabrication
minimizes the influence of the environment, since the dots are
formed under high vacuum in a cryostat (keeping contamination to a
minimum) and the dots are suspended (reducing electrostatic
interaction with the substrate). In the future this technique might
be improved to enable the control of the dot size (perhaps down to a
single atom). For this, a conductance monitoring system operable on
a much faster time scale is required. In addition, working in a
cleaner gas environment may allow to better control the chemistry at
the edges. For instance, the rupture of the sheet could be carried
out in ultra-pure hydrogen in order to terminate carbon atoms with
hydrogen atoms. Our suspended graphene quantum dots are potentially
interesting nanoelectromechanical systems. Suspended graphene can
act as a mechanical resonator whose vibrations could be coupled to
charge transport through the quantum dot, as demonstrated in other
materials \cite{weig}.

We are grateful to A. Barreiro for help with fabrication and
acknowledge fruitful discussions with J. G\"{u}ttinger. This work
was supported by an EURYI grant and FP6-IST- 021285-2.


\begin{references}
\bibitem{note}
The stability diagrams of some of our devices revealed a series of
dots, yet they were not significantly altered by subsequent high
current treatments.
\bibitem{thermal}
The analysis of the thermal linewidth of conductance resonances at
low $V$ leads to the lever arm conversion between $V_{g}$ and the
actual change in the electrochemical potential, which is consistent
with the slope of the diamond in Fig.~2(b).
\end{references}
\end{document}